\documentstyle[multicol,aps,prb,epsf]{revtex}

\newcommand{\Vec}[1]{\mbox{\boldmath$#1$}}

\begin{document}

\draft

\title{Wrapping Current versus Bulk Integer Quantum Hall Effect in 
Three Dimensions}
\author{Mikito Koshino, Hideo Aoki}
\address{Department of Physics, University of Tokyo, Hongo, Tokyo
113-0033, Japan}
\author{Bertrand I. Halperin}
\address{Physics Department, Harvard University, Cambridge MA 02138}
\date{\today}

\maketitle

\begin{abstract}
Surface electron currents are studied for the integer 
quantum Hall effect in three dimensions(3D) proposed 
previously by Kohmoto et al. and by Koshino et al.   
We predict the current wraps the facets of the sample 
with its intensity and direction dictated by 3D Chern numbers 
which are just the quantized Hall conductivities in 3D 
($\sigma_{xy}, \sigma_{zx}$), 
so a natural connection exists between the surface and bulk currents just 
as in 2D.  An experiment to detect the 3D integer quantum Hall effect 
through the wrapping current is proposed.
\end{abstract}

%\newpage

\begin{multicols}{2}
\narrowtext

%\section{Introduction}

The chiral edge states of 2D electrons in magnetic fields have been
extensively investigated for the quantum Hall effect(QHE)\cite{Halp1982}. 
Namely, edge states exist for each Landau gap in a finite quantum Hall 
system, and the Hall current carried by them is shown to 
coincide exactly with one calculated with the Kubo formula\cite{AokiAndo} 
for the bulk sample, which has been interpreted 
in terms of topological quantum numbers characterizing 
the quantum Hall currents\cite{Hats}.

While the QHE is usually conceived as specific to two-dimensional 
systems, it is known that integer QHE can occur even in three dimensions (3D)  
if the spectrum has an energy gap and if the Fermi energy 
lies in the gap\cite{Halp,Mont,Kohm}.  
Although gaps do not usually appear in 3D, two of the present authors 
have shown\cite{Kosh} that it is possible to
have a class of energy spectra having a series of gaps (Hofstadter butterfly) 
in 3D lattice systems in an appropriate condition, 
where we have a 3D-specific IQHE, i.e., each of the Hall conductivities 
$\sigma_{xy}, \sigma_{zx}$ quantized for each gap.  
That analysis for the 3D QHE is based on the bulk description, 
and a natural question we can now ask is: whether and how 
surface states appear in this 3D QHE.   In superlattice systems, i.e., stack of 
2D systems, the surface current has been intensively studied.\cite{Bale} 
There, one discusses a stack of the chiral edge states, 
called the chiral sheath current.  
Surface states are also discussed for 
the field-induced spin density waves (FISDW) in organic 
conductors\cite{Yako,Seng}. In those systems, however,
the inter-layer hoppings are relatively so small that 
the surface states may be understood as weakly coupled 
2D edge states.  By contrast, here we are talking 
about the 3D-specific IQHE with $\sigma_{xy}, \sigma_{zx}$ quantized 
as 3D Chern (topological) numbers.  

So we have studied here the surface states for 3D in general and
for the 3D butterfly in particular.  We show that there exists a 
surface {\it wrapping current} that winds around the facets of the 3D
sample.  A hallmark of the 3D-specific nature appears as 
the current on each facet flowing 
obliquely to the crystallographic axes, 
where the current direction is given in terms of
the 3D Chern numbers.  An interesting observation 
is that the 3D Hall currents carried by surface states are exactly the same as
what is given by the bulk conductivities.

%\section{Thermodynamic argument for the wrapping current}

\begin{figure}
\begin{center}
  \leavevmode\epsfxsize=80mm \epsfbox{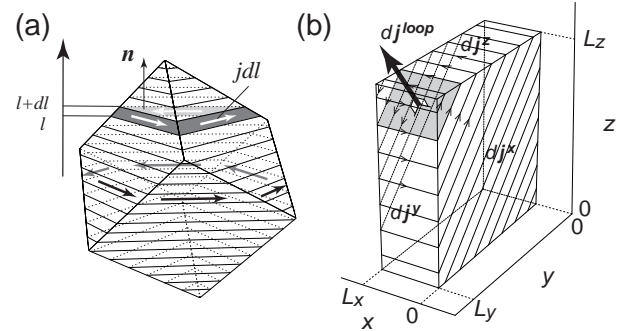}
\end{center}
\caption{(a)A schematic picture of the surface wrapping current.  
$\Vec{j}^{\rm loop}=j\Vec{n}$ is the surface loop current density.
(b)The wrapping current calculated for the microscopic model for
the 3D QHE case (eqs.(\ref{j_yz}) and (\ref{j_zx})).
}
\label{fig_wrapcurrent}
\end{figure}

Although we can expect a 3D sample in a magnetic field spontaneously 
carries a surface current in equilibrium as a 2D sample carries an 
edge current, the situation should be 
more complex, since the surface current will have to wind around 
various faces of the 3D sample.   
However, we can show, from thermodynamics, that the surface current may be 
expressed in a surprisingly simple way in terms of the bulk density of 
states. Let us take a cubic, homogeneous sample 
for simplicity and put it in a uniform magnetic field $\Vec{B}$.
We assume that the surface current is uniform over 
each face of the cube and that the effect of 
cube edges can be neglected.  
Then we can characterize the surface currrent flowing 
around the faces of the cube by regarding it as a 
bunch of loop-current segments as depicted in Fig.\ref{fig_wrapcurrent}(a).  
Since the current is assumed to be uniform on each face,
the bunch is characterized by only two quantities: 
the vector $\Vec{n}$, normal to each loop, 
and $j$, the loop current intensity flowing within 
a unit height measured along $\Vec{n}$.
If we define the loop current-density vector as
$ \Vec{j}^{\rm loop} \equiv j \Vec{n}$, 
the magnetization associated with the current is 
$ \Vec{m} = V \Vec{j}^{\rm loop}$, where $V$ is the sample volume.
Combining with a thermodynamic Maxwell's relation,
we obtain 
\begin{equation}
 \frac{d \Vec{j}^{\rm loop}}{\partial \mu} =  
 \frac{\partial \rho}{\partial\Vec{B}},
 \label{dj_dmu}
\end{equation}
where $\rho$ is the density of occupied states per unit volume and
$\mu$ the chemical potential.  

Now if we move on to the QHE
case in the 3D periodic system, by assuming
that the Fermi energy $E_F$ lies in a gap of 
the bulk energy spectrum\cite{Mont,Kohm}, 
we have a relation
\begin{equation}
 \frac{\partial\rho}{\partial \Vec{B}} = -\frac{e}{2\pi h} \Vec{J},
 \label{J}
\end{equation}
where $-e$ is the charge of an electron and $\Vec{J}$ is a 
reciprocal vector of the periodic potential. 
$\Vec{J}$ depends on the gap in which $E_F$
lies and gives the quantized 3D Hall conductance through
\begin{equation}
\sigma^{\rm bulk}_{ij} = \frac{e^2}{2\pi h} \sum_{k} \epsilon_{ijk} J_k, \nonumber
\end{equation}
where $\epsilon_{ijk}$ is the unit antisymmetric tensor.
From eqs.(\ref{dj_dmu}) and (\ref{J}),
the surface loop current carried by the surface
states between $E$ and $E+d\mu$ is given as
\begin{equation}
 d\Vec{j}^{\rm loop} =  -\frac{e}{2\pi h} \Vec{J} d\mu.
\label{j_loop}
\end{equation}
From this we will conclude that there is a quantized wrapping
current. Even at this stage, we can see a unique behavior: that the
direction (in real space) of the surface current should always be
perpendicular to a fixed reciprocal vector, which in general deviates
from the symmetry axes of the crystal and also from $\Vec{B}$, but which is
solely determined by the gap in which $E_F$ is located.

%\section{Microscopic picture of the wrapping currents}

\begin{figure}
\begin{center}
  \leavevmode\epsfxsize=70mm \epsfbox{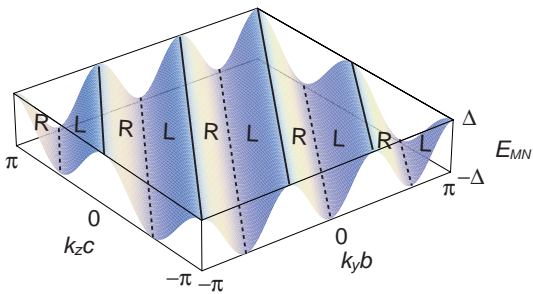}
\end{center}
\caption{The energy dispersion for the surface states on the
 $\hat{\Vec{x}}$ faces in the energy gap $(M,N)=(3,2)$. 
 `L' and `R' represent the surface states 
on the left ($x=0$) and the right ($x=L_x$) faces, respectively. 
}
\label{fig_dispersion}
\end{figure}

Having looked at the thermodynamics, let us 
move on to a microscopic model for the 3D QHE surface currents.
We take a noninteracting tight-binding
electron system in a uniform magnetic field $\Vec{B}$.
Schr\"{o}dinger's equation is 
$-\sum_j t_{ij} e^{i\theta_{ij}}\psi_j = E\psi_i$. 
Here $\psi_i$ is the wave function at site $i$,
the summation is over nearest-neighbor sites, and 
$\theta_{ij} = \frac{e}{\hbar}\int_{j}^{i}\Vec{A}\cdot {\rm d}\Vec{l}$ 
is the Peierls phase factor arising from the magnetic flux 
where \Vec{A} is the vector potential with $\nabla\times\Vec{A} = \Vec{B}$. 
We consider a 3D simple-cubic lattice with lattice constants $a,b,c$ 
and $\Vec{B}=(0,B\sin\theta ,B\cos\theta)$ lying in the $yz$ plane.
In the gauge $\Vec{A}=(0,Bx\cos\theta,-Bx\sin\theta)$, 
$y,z$ become cyclic, and we have 
$\psi_{lmn} = e^{i k_y m b + i k_z n c} F_l$, 
where $l,m,n$ are the site indexes along $x,y,z$, respectively.
Schr\"{o}dinger's equation then reads
\begin{eqnarray}
  -t_x(F_{l-1} \,+\, F_{l+1}) \,-\,  [2t_y \cos(G_b l a
  +k_y b) \nonumber \\
  + 2t_z \cos(-G_c l a +k_z c)] F_l = E F_l
  \label{Harp3D}.
\end{eqnarray}
Here $t_x, t_y, t_z$ are the transfer integrals between nearest
neighbors along $x,y,z$, respectively, and 
$(G_b,G_c) = \frac{e}{\hbar} (B_z b, B_y c)$.

Now we consider that the system is quasi-1D, i.e., $t_x \gg t_y,t_z$,
which is required for the 3D butterfly\cite{Kosh}.
By applying the effective-mass approach for the $x$ direction, we have
\begin{eqnarray}
 \bigl[ E_x(\partial_x)
  &-& 2t_y \cos(G_b x + k_y b) \bigr. \nonumber \\
  \bigl. &-& 2t_z \cos(-G_c x +k_z c)\bigr] F(x) = EF(x),
\end{eqnarray}
where $E_x(\partial_x) = (a t_x \partial_x)^2$.
The perturbation ($\propto t_y, t_z$) mixes
eigenfunctions $\{F(x) = e^{ik_x x}\}$,
and energy gaps of magnitude $\Delta =|t_y|^M|t_z|^N$ 
(in a unit of energy $t_x = 1$) open at 
$k_x = \pm \frac{1}{2}(MG_b + NG_c),$
where $M,N$ are integers.
In the quasi-1D limit $t_y,t_z \rightarrow 0$, 
Schr\"{o}dinger's equation around the gap $(M,N)$
simplifies into
\begin{equation}
 \left(
  \begin{array}{cc}
   -i\hbar v \partial_x & \Delta e^{i\varphi}\\
   \Delta e^{-i\varphi}   &   i\hbar v \partial_x 
  \end{array}
 \right)
 \left(
  \begin{array}{c}
   F_+(x)\\   F_-(x)
  \end{array}
 \right)
 = E
 \left(
  \begin{array}{c}
   F_+(x)\\   F_-(x)
  \end{array}
 \right),
 \label{Schr}
\end{equation}
where we have decomposed the wave function 
into the left- and right-moving components, $F = F_+ + F_-$, 
around $k_x = \pm k_{\rm F} = \pm \frac{1}{2}(MG_b + NG_c)$, 
$\varphi = M k_y b - N k_z c$, and $v = \hbar k_{\rm F}/m^*$.

Now we consider the surface states at the gap $(M,N)$
by extending the discussion for the 
FISDW\cite{Yako,Seng}, 
which is one way of realizing the energy gaps in 3D systems 
and also mathematically similar to 
the 3D QHE considered in ref.\cite{Kosh,KoshFISDW}.
We first consider the surfaces normal to the conductive axis 
($\hat{\Vec{x}}$) in a sample with $0 \leq x \leq L_x$.  
By applying the boundary condition $F(0) = 0$ to 
the differential eq.(\ref{Schr}),
we find a surface state that decays into the bulk,
\begin{equation}
 F(x) = e^{-\kappa x} \sin k_{\rm F} x, 
\end{equation}
where $\kappa = (\Delta/\hbar v) \sin (M k_y b - N k_z c)$.
The corresponding eigenenergy measured from the gap center is
\begin{equation}
 E_{MN} = -\Delta \cos (M k_y b - N k_z c).
\label{E_MN}
\end{equation}
The solution with $\kappa > 0$ corresponds to the left surface
($x\simeq 0$) while $\kappa < 0$ to the right ($x\simeq L_x$).  The
energy dispersion of the surface state must be plotted 
against $(k_y, k_z)$ as in 
Fig.\ref{fig_dispersion}. They oscillate $M(N)$ times 
along $k_x(k_y)$, where we
can see the stripe-like areas alternating for the right and left surface 
states.  Since the Brillouin zone is topologically a torus, we can
define two winding numbers for the stripe along toroidal and poloidal
directions.  They are just $M$ and $N$, and these two numbers
are in fact the Chern numbers in 3D, which correspond to 
$\sigma_{xy}, \sigma_{zx}$ 
as shown below.  For 2D Hatsugai has shown that the edge
states whose energy dispersions correspond to wiggly lines against $k_y$
between the Landau subbands have topological numbers.\cite{Hats} Hence
the present result is a natural extension to 3D.  Although small 
$t_y, t_z$ are assumed above, the 3D Chern numbers, being topological, 
should be constant for larger $t_y,t_z$.

From this microscopic model, we can actually 
calculate the current density on the
surface.  The expected value of the velocity for the electron for given
$(k_y,k_z)$ is
\begin{equation}
 \Vec{v} = \frac{1}{\hbar} \frac{\partial E_{MN}}{\partial \Vec{k}}
  = \pm \frac{1}{\hbar} \sqrt{\Delta^2 - E_{MN}^2}(0, Mb, -Nc),
\label{vec_v}
\end{equation}
where $+$ and $-$
corresponds to the left and right surface states, respectively.
We can see that every state in the gap $(M,N)$ has a velocity
parallel to a single vector $(Mb, -Nc)$.
The current density on each surface 
carried by the states between $E$ and $E+d\mu$ is then expressed as
\begin{equation}
 d\Vec{j}^{\hat{\Vec{x}}}
= \mp\frac{e}{h}\left(0, \frac{M}{c}, -\frac{N}{b}\right)d\mu
\, \, \, ({\rm at} \, \, x =0, L_x).
\label{j_yz}
\end{equation}

The derivation of the surface currents on the planes 
$\perp \hat{\Vec{y}}$ or $\hat{\Vec{z}}$ that contain 
the conductive axis is slightly different.  
We consider a finite sample with 
$0 \leq y \leq L_y, 0 \leq z \leq L_z$.  
To see which states are mixed, we can write 
the perturbational term (the off-diagonal term in eq.(\ref{Schr})) 
for the gap $(M,N)$
in a second quantized form in $k$ space as
\begin{eqnarray}
 {\cal H}' &=& \sum_{k_x,k_y,k_z} 
  \Delta e^{i(Mk_yb - Nk_zc)} \nonumber \\
 && \times c^{\dagger}_{k_x + MG_b + NG_c, k_y, k_z}c_{k_x, k_y, k_z} 
  + {\rm H.c.},
\end{eqnarray}
where $c^{\dagger} (c)$ are creation (annihilation) opperators.  
In the Wannier representation for $y,z$, this becomes
\begin{eqnarray}
  {\cal H}' = \sum_{k_x,m,n} \Delta 
  c^{\dagger}_{k_x + MG_b + NG_c, m-M, n+N}c_{k_x, m, n} + {\rm H.c.},
\end{eqnarray}
where $|k_x, m ,n\rangle$ is a mixed Wannier-Bloch 
basis localized at $(y,z) = (mb,nc)$ 
and delocalized along $x$ (so we call it a `chain'). 
Once ${\cal H'}$ is switched on,
the states at $(m,n)$ with $k_x > 0$ and $(m-M,n+N)$ 
with $k_x < 0$ are mixed, and an energy gap appears at 
$k_x = \pm \frac{1}{2}(MG_b + NG_c)$.
We can immediately see that chains lying 
within $M(N)$ lattice constants of the faces 
$\perp \hat{\Vec{y}}(\hat{\Vec{z}})$ do not couple to other chains,
so these states remain gapless, which is exactly 
the origin of the surface states.  
The expectation value of the velocity along $x$ is
equal to $\pm \frac{\hbar}{2m^*}(MG _b + NG_c)$
with opposite directions between the two sides.
The current density for each surface for energies between $E$ and $E+d\mu$ 
is
\begin{eqnarray}
 d\Vec{j}^{\hat{\Vec{y}}} &=& \pm\frac{e}{h}\left(\frac{M}{c},0,0\right)d\mu
\, \, \, ({\rm at} \, \, y =0, L_y), \nonumber\\
 d\Vec{j}^{\hat{\Vec{z}}} &=& \mp\frac{e}{h}\left(\frac{N}{b} ,0,0\right)d\mu
\, \, \, ({\rm at} \, \, z =0, L_z).
\label{j_zx}
\end{eqnarray}
From eqs.(\ref{j_yz}),(\ref{j_zx}), 
we can see that the currents satisfy Kirchhoff's law
on each edge of the sample, so that we end up with a current sheet 
that {\it wraps} the whole surface (Fig.\ref{fig_wrapcurrent}(b)), and  
the corresponding loop current segment defined above is 
$d\Vec{j}^{\rm loop} = 
\frac{e}{h}\left(0, \frac{N}{b}, \frac{M}{c}\right)d\mu$.  
We can see that this result is consistent with the formula(\ref{j_loop}), 
since the reciprocal vector $\Vec{J}$ (eq.(\ref{J})), 
calculated originally by Montambaux and Kohmoto\cite{Mont}, is
$(0, -2\pi N/b, -2\pi M/c)$  for the present case.  

%\section{Surface vs bulk currents and experimental setup}

\begin{figure}
\begin{center}
  \leavevmode\epsfxsize=80mm \epsfbox{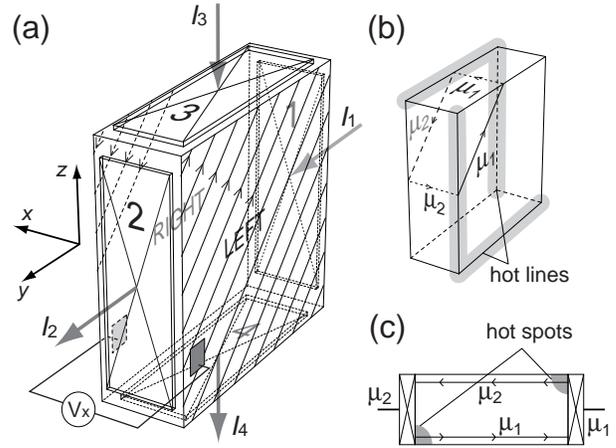}
\end{center}
\caption{
(a)The experimental setup to detect 3D QHE 
surface currents.
The arrows represent the direction of the currents,
which is opposite to the motion of electrons.
(b)The ``hot line'' in the 3D QHE experiment 
with (c)the corresponding picture in 2D.}
\label{fig_experiment}
\end{figure}

In the 3D QHE system, how to measure the conductivity tensor experimentally
is even less trivial than in 2D.  We propose here an experiment to
detect 3D QHE surface currents in analogy with the
2D Hall bar experiment. 
As shown in Fig.\ref{fig_experiment} we attach two pairs of electrodes, 
$(1,2)\perp \hat{\Vec{y}}$ and $(3,4)\perp \hat{\Vec{z}}$.   
The incoming or outgoing currents, 
denoted as $I_1,\cdots,I_4$ as depicted in the figure, 
are a function of 
the chemical potentials of the electrodes, $\mu_1,\cdots,\mu_4$.  
We can calculate the currents by first assuming that 
all the currents are carried by the surface states.  For the 
present electrodes, 
we are left with the surface currents on the two 
$\hat{\Vec{x}}$ faces (which we call right and left 
surfaces), whose current density is given in eq.(\ref{j_yz}).  

As for the chemical potential, 
let us consider the situation where all $\mu$'s are in the gap
$(M,N)$ with $M,N>0$.  Then we can see from eq.(\ref{j_yz}) that the
electrons on the left surface flow from the electrode 1 or 3 (source) 
into 2 or 4 (drain) with a reverse current for the right surface. 
So when we set
$\mu_1 = \mu_3 \neq \mu_2= \mu_4$, 
each $\hat{\Vec{x}}$ face should be in
equilibrium with the source electrode 
($\mu_1$ for the left surface, $\mu_2$ for the right) 
if we neglect dissipations in the sample 
(which we will touch upon later).  
The net currents must be conserved, which implies, 
for rectangular surfaces, 
$I_1 = I_2 (\equiv I_y)$ and $I_3 = I_4 (\equiv
-I_z)$, and they are readily calculated, via eq.(\ref{j_yz}), as
\begin{equation}
 (I_y, I_z) = \frac{e}{h}\left(-\frac{L_zM}{c}, \frac{L_yN}{b}\right)
 (\mu_1-\mu_2).
\end{equation}
The Hall conductivity tensor due to the surface conduction becomes
\begin{equation}
 (\sigma^{\rm surface}_{xy} , \sigma^{\rm surface}_{zx})
= -\frac{e^2}{h}\left(\frac{M}{c}, \frac{N}{b}\right),
\end{equation}
where we have put $\mu_1-\mu_2 = -eV_x$ 
with $V_x$ being the Hall voltage.
If we compare this with the expression for the bulk conductivity\cite{Mont},
we can establish a relationship for the Hall conductivities, 
$\sigma_{ij}^{\rm surface} =\sigma_{ij}^{\rm bulk}$ 
in 3D, i.e., the result {\it does not change
whether the currents flow in the bulk or on the surface}.
In the usual 2D QHE, Hatsugai\cite{Hats} shows that 
$\sigma_{xy}^{\rm surface} =\sigma_{xy}^{\rm bulk}$ 
by identifing the connection between the topological integers
for the bulk and the edge states.
So the discussion here in terms of the 3D topological numbers shows 
that this property remarkably extends to 3D.

% Another way to see why sigma_bulk = sigma_edge in 3D

\begin{figure}
\begin{center}
  \leavevmode\epsfxsize=60mm \epsfbox{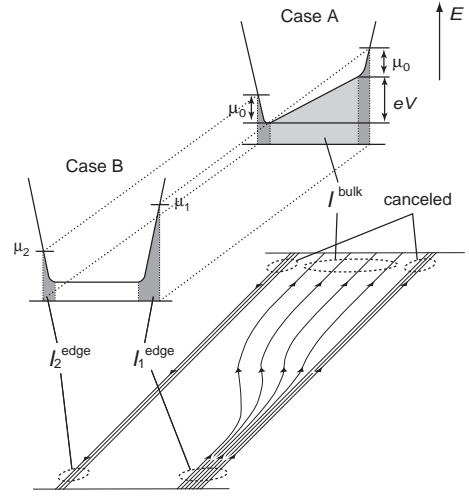}
\end{center}
\caption{An intuitive picture which explains
the correspondence $\sigma^{\rm surface}$ and 
$\sigma^{\rm bulk}$ in the 2D Hall bar.
}
\label{fig_edge_bulk}
\end{figure}

We can give an intuitive 
way to understand why surface or bulk does not really matter.   
Let us start with 2D QHE to consider two possible 
situations (Fig.\ref{fig_edge_bulk}): 
Case A has an electrostatic-potential gradient over the bulk with 
the potential drop, $eV$, assumed to be the same as 
the chemical potential difference across the two edges.  
Case B has no potential drop in the bulk while 
the chemical potentials at two edges,
$\mu_1,\mu_2$, differ.  
Physically, the situation is determined by, e.g., 
how the local equilibrium is achieved by inelastic 
processes as studied by Ando in a numerical 
calucation of the potential profile\cite{Ando}.
We can now question how the $\sigma$ 's in two pictures are connected.
If we put $\mu_1-\mu_2 = eV$, 
we can envisage that the two cases cross over to 
each other continuously as in Fig.\ref{fig_edge_bulk}.
The only assumption is that 
we can neglect the scattering across the edge states on the two sides.
The current in case A can be divided into the bulk current $I^{\rm bulk}$
and edge currents\cite{Komi}. The edge currents on the left and right edges
have opposite directions but the same intensity because 
the chemical potential at either edge is assumed to have 
the same energy difference (denoted by $\mu_0$ in the figure) 
from the Landau level center.  
In case B, the conduction is entirely due to the edge currents
$I_1^{\rm edge}, I_2^{\rm edge}$, which are different 
because $\mu_1 \neq \mu_2$.
Since the total current must be preserved 
as we go from case A to B, we have
$ I_1^{\rm edge} - I_2^{\rm edge} = I^{\rm bulk}$,
which is in fact the equality in question.   
If we regard the 3D system 
as a stack of current segments (or 2D Hall "bars") 
as in Fig.\ref{fig_experiment}(b), 
this argument can be extended, 
since currents across the adjacent "bars" are absent. 
Then we reproduces the property 
$\sigma^{\rm surface} =\sigma^{\rm bulk}$.  

As a final comment, 
in the 2D Hall bar geometry it has long been recognized that 
there are two hot spots 
where the chemical potential has to drastically drop from $\mu_1$ to $\mu_2$
dissipatively. In our 3D geometry, the hot spots should 
become two ``hot lines'' as shown in Fig.\ref{fig_experiment}(b).
Given that the required magnetic field ($< 40$ T) is well within 
experimental feasibility as estimated in \cite{Kosh,KoshFISDW}, 
experimental detection of the wrapping currents and hot lines 
should be interesting.

%Acknowledgement
H.A. wishes to thank Sinya Uji for helpful discussions.  M.K. would like to
thank the JSPS Research Fellowships for Young Scientists for financial
support.  B.I.H. acknowledges support by NSF grant DMR 99-81283.

\end{multicols}

\begin{thebibliography}{99}
\bibitem{Halp1982}
B. I. Halperin, Phys. Rev. B {\bf 25} 2185 (1982)
\bibitem{AokiAndo} H.  Aoki  and T. Ando, Solid State Commun. 
{\bf 38}, 1079 (1981).
\bibitem{Hats}
Y. Hatsugai, Phys. Rev. Lett. {\bf 71} 3697 (1993).
\bibitem{Halp}
B. I. Halperin, Jpn. J. Appl. Phys. Suppl. {\bf 26}, 1913 (1987).
\bibitem{Mont}
G. Montambaux and M. Kohmoto, Phys. Rev. B {\bf 41}, 11417 (1990).
\bibitem{Kohm}
M. Kohmoto, B. I. Halperin, and Y. Wu, Phys. Rev. B {\bf 45}, 13488 (1992).
\bibitem{Kosh}
M. Koshino, H. Aoki, K. Kuroki, S. Kagoshima, and T. Osada,
Phys. Rev. Lett. {\bf 86}, 1062 (2001); Phys. Rev. B {\bf 65}, 045310 (2002).
\bibitem{Bale}
L. Balents and M. P. A. Fisher, Phys. Rev. Lett. {\bf 76}, 2782 (1996). 
\bibitem{Yako}
V. M. Yakovenko, and H.-S. Goan,
J. Phys. (France) I {\bf 6}, 1917 (1996).
\bibitem{Seng}
K. Sengupta, H.-J. Kwon, and V. M. Yakovenko,
Phys. Rev. Lett. {\bf 86}, 1094 (2001).
\bibitem{KoshFISDW} 
M. Koshino, H. Aoki, and T. Osada, Phys. Rev. B, to be published.
%\bibitem{Wido}
%A. Widom, Phys. Lett. {\bf 90A}, 474 (1982).
%\bibitem{Stre}
%P. St\v{r}eda, J. Phys. C {\bf 15}, L718 (1982).
\bibitem{Ando}
T. Ando, Surface Science {\bf 361/362}, 270 (1996).
\bibitem{Komi}
S. Komiyama and H. Hirai, Phys. Rev. B {\bf 54}, 2067 (1996).
\end{thebibliography}
\end{document}